\documentclass[12pt,preprint]{aastex}
\usepackage{natbib}

\shorttitle{Eccentric orbits can hide resonant planets}
\shortauthors{Anglada-Escud\'e, L\'opez-Morales \& Chambers}

\begin{document}

\title{How eccentric orbital solutions can hide planetary systems in 
2:1 resonant orbits}

\author{
Guillem Anglada-Escud\'e\altaffilmark{1}, 
Mercedes L\'opez-Morales\altaffilmark{1,2}, 
John E. Chambers\altaffilmark{1}}

\email{
anglada@dtm.ciw.edu, 
mercedes@dtm.ciw.edu, 
chambers@dtm.ciw.edu}

\altaffiltext{1}{Carnegie Institution of Washington, Department of Terrestrial 
Magnetism, 5241 Broad Branch Rd. NW, Washington D.C., 20015, USA}
\altaffiltext{2}{Hubble Fellow}

\begin{abstract} 

The Doppler technique measures the reflex radial motion of a star
induced by the presence of companions and is the most successful
method to detect exoplanets. If several planets are present, their
signals will appear combined in the radial motion of the star,
leading to potential misinterpretations of the data. Specifically,
two planets in 2:1 resonant orbits can mimic the signal of a
single planet in an eccentric orbit. We quantify the implications
of this statistical degeneracy for a representative sample of the
reported single exoplanets with available datasets, finding that 
1) around $35\%$ percent of the published eccentric one-planet
solutions are statistically indistinguishible from planetary
systems in 2:1 orbital resonance, 2) another $40\%$ cannot be
statistically distinguished from a circular orbital solution and
3) planets with masses comparable to Earth could be hidden in
known orbital solutions of eccentric super-Earths and Neptune mass
planets.

\end{abstract}


\section*{Introduction}

Most of the $+300$ exoplanets found to date have been
discovered using the Doppler technique, which measures the
reflex motion of the host star induced by the planets
\citep{mayor:1995,marcy:1996}. The diverse characteristics
of these exoplanets are somewhat surprising. Many of them are
similar in mass to Jupiter, but orbit much closer to their host
stars. This finding has led to extensive work on planet
formation and migration theories to explain how those planets
got to their present location \citep[eg.][]{Ward:1997,Ida:2004}. Also,
many of the planets seem to move in orbits with eccentricies
significantly larger than those observed in the Solar System
(where $e<0.1$, except for Mercury which has $e\sim0.2$), 
This result poses a problem for the planet formation theories of
 Core Accretion \citep{pollack:1996} and Disk Instability
\citep{boss:1997}, since both predict that planets form in
quasi--circular orbits. The current explanation is that large
eccentricities are triggered by secular interactions, i.e. the
{\it Kozai effect} \citep{soder:1975}, or by rare close-in
encounters \citep{ford:2005}. The true distribution of exoplanet
eccentricities is therefore key to understand the formation of
planetary systems \citep{thommes:2008}.

About $25$\% of the planets detected to date are in multiplanetary systems.
The first multi-planet system was discovered around the solar type
star $\upsilon$ Andromedae \citep{butler:1997}. Currently we know
of three planets in that system with masses between 0.69 and 3.95
$m_{Jup}$ and orbital periods between 4.6 and 1275 days. The second
multi-planet system discovery was around the M dwarf Gliese 876
\citep{delfosse:1998}. In that case, the first
detected planet (Gl 876b, with Msini = 1.935 $M_{Jup}$ and P =
60.94 days), was the one orbiting furthest from the star. A better
sampling of the Doppler curve led to the subsequent discovery of
the other two planets, Gl 876c, a 0.56 $M_{Jup}$ planet
orbiting with a period  of 30.1 days and Gl 876d, a closer-in 0.018
$M_{Jup}$ planet with a period of only 1.94 days. Gl 876b
and c, in particular, are in 2:1 resonance \citep{laughlin:2001}. 

Interestingly, multi-planet systems are usually first detected as a
single planet in a significanly eccentric orbit. Then, every time an additional
planet is found, the eccentricities of the already known
planets tend to decrease. This is because spurious harmonics due to
random noise and uneven sampling are absorbed at half of the period of
the detected planet when an eccentric solution is forced
\citep[see][for a more detailed discussion on  spurious
eccentricities]{lucy:2005}. 55 Cnc \citep{fischer:2008} and 
HD160691\citep{pepe:2007} are the most clear examples.
Both planetary systems started with detections of single massive planets with
relatively large eccentricities (i.e. $e>0.3$). At present, the four
planets detected in HD160691 and the five planets in 55 Cnc have all
eccentricities smaller than $0.2$.  Table \ref{55cnc} illustrates the
evolution of the published eccentricities in 55 Cnc as more and more
planets were discovered.

In this paper we explore the case where two--planet systems can be
confused with single planets in eccentric orbits. The situation
arises  when two planets are in a circular 2:1 mean motion resonance
being the outer one the most massive ($m_{in}/m_{out}\le 0.5$). Since
a full Keplerian solution for a single planet is the natural
choice, the statistical degeneracy explained in detail in
Sec.~\ref{sec:math} introduces an observational bias towards eccentric
solutions. In the forthcoming sections we show how this bias can have 
important implications for the known sample of extrasolar planets,
that is, a significant fraction of the reported eccentric planets may
in fact be multiple systems in nearly circular orbits, 
and several very low mass planets might have been already detected,
but their effect on the star has been misinterpreted as an orbital
eccentricity of an outer more massive planet.

\section{Mathematical solution degeneracy}\label{sec:math}

The solution degeneracy between a single planet eccentric orbit and two
planets in circular resonant orbits is a direct consequence of the well
known Fourier expansion of the Kepler equation into powers of the 
eccentricity \citep[see][as an example]{moulton:1914}. In
\cite{konacki:1996}, the method of frequency analysis was first applied to
an extrasolar planetary system and \cite{konacki:1999} adapted it to
Doppler measurements. The potential confusion between eccentric orbits and
resonant systems has been briefly mentioned in \citep{marcy:2001} and
\cite{ford:2006}, but this issue has not been specifically considered
until now in a broad statistical sense.

Mathematically, the degeneracy between the resonant and the eccentric
solutions comes from the fact that their equations of Keplerian
trajectories are identical up to first order in the eccentricity.
Detailed analytical expressions in terms of the Bessel functions can
be found elsewhere \citep[eg.][]{konacki:1999}. The relevant terms up
to the 7th power in the eccentricity can be found in
\citet{lucy:2005}. Here we only discuss the first order term, which is
the one relevant to the degeneracy under discussion

\begin{figure}[htb]
\center
\includegraphics[width=6in]{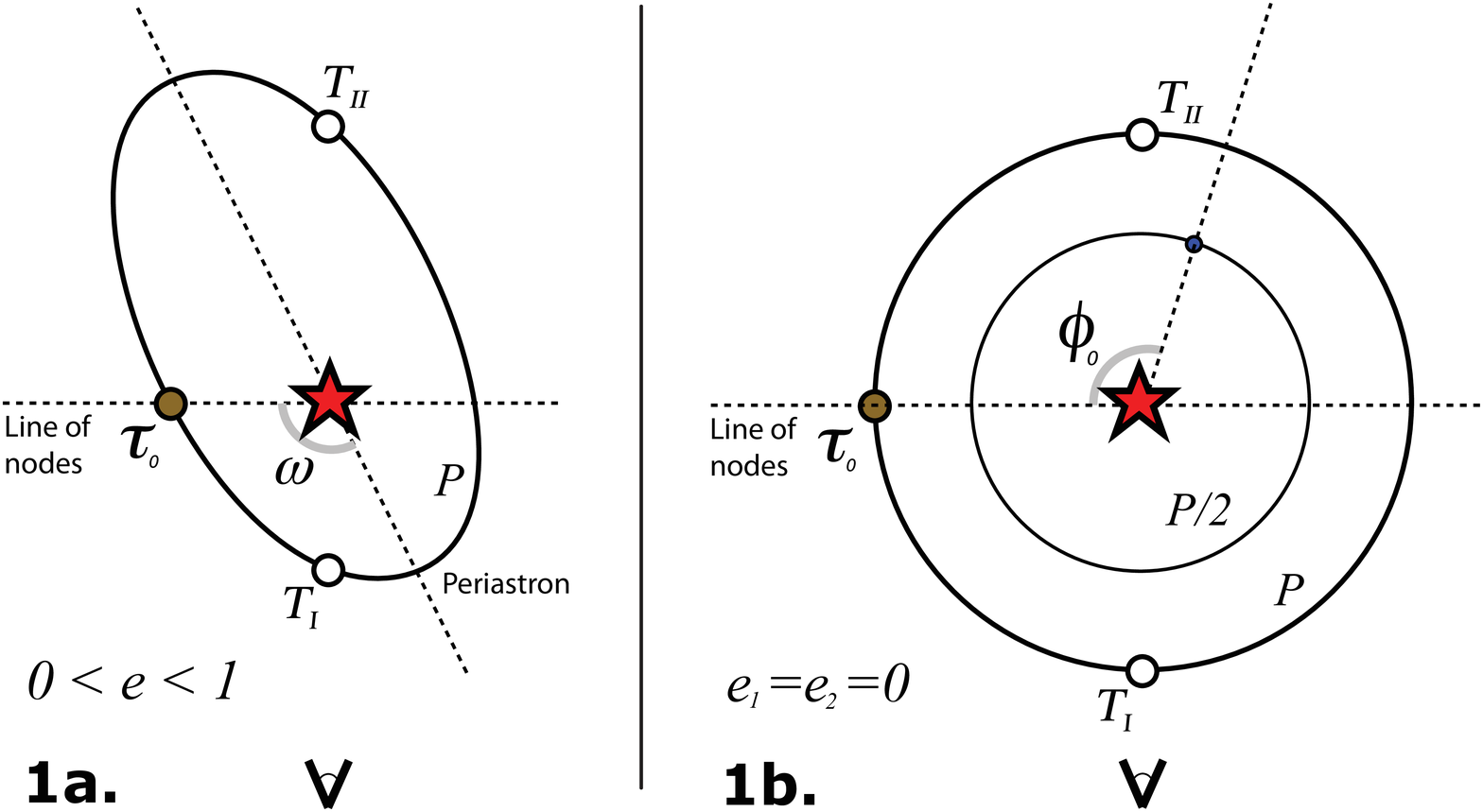}
\caption{Seen from \textit{above}, diagrams of
the relevant orbital parameters of one planet in an eccentric
orbit (\textit{Left}), and two planets in a 2:1 resonant
circular orbit (\textit{Right}). $\tau_0$ is the instant of
crossing of the line of nodes. The instants of transit and
occulation are marked as $T_I$ and $T_{II}$.}\label{fig:dibu}
\end{figure}

In the case of a single eccentric planet, the reflex radial velocity
motion of the star is
\begin{eqnarray}
v^e_r = v_{r0} &+& K \cos \left[ W\left(t-\tau_0\right) \right] \, 
\label{eq:velecc} \\
      &+& 
Ke\cos \left[ 2 W\left(t-\tau_0\right) - \omega\right] \nonumber \\
&+& O({Ke^2})\,, \nonumber
\end{eqnarray}
\noindent where $v_{r0}$ is the linear radial velocity of the
barycenter of the system, $K$ is the semi-amplitude of the radial
velocity variations induced by the planet on the star, $\omega$ is the argument of
the periastron (angle between the periastron of the orbit and the
ascending node), $\tau_0$ is the time of crossing of the ascending
node, and $W=2\pi/P$ is the orbital frequency, where $P$ is the
orbital period (see Fig.~\ref{fig:dibu}a). The term proportional to
$Ke$ is called the \textit{first eccentric harmonic}, while the term
$O({Ke^2})$ contains all the higher order contributions. 

If instead we have a two--planet system, both in circular orbits and
the inner planet having an orbital period half of the outer one,
i.e. $W_2=2W$ (see Fig.~\ref{fig:dibu}b), the expression for
the radial velocity of the star is
\begin{eqnarray}
v^R_r = v_{r0} 
&+& k_1 \cos \left[ W\left(t-\tau_0\right) \right] \label{eq:velres}\\
&+& k_2 \cos \left[ 2 W\left(t-\tau_0\right) + \phi_0\right]\nonumber\, \\
&+& O({k_1e_1,k_2e_2, Ke^2}),\nonumber
\end{eqnarray}
\noindent where $k_1$ and $k_2$ are the radial velocity
semi--amplitudes of the outer and the inner planet. $W$ and
$\tau_0$ are the orbital frequency and the time of crossing of the
ascending node of the outer planet, and the angle $\phi_0$ is the
relative phase between the two planets at $\tau_0$. Higher order
terms, summarized here as $O({k_1e_1,k_2e_2, Ke^2})$, become
significant if the orbits are allowed to be eccentric.

To a first order approximation, $v^e_r$ and $v^R_r$ are formally
identical if $k_1 = K$, $k_2 = Ke$, and $\phi_0 = -\omega$. This
implies that the signal $k_2$ of an inner lower-mass planet will be
indistinguishable from the \textit{first eccentric harmonic} $Ke$
unless the observations are precise enough to resolve the second order
term in the harmonic expansion. The amplitude of that second order
term is $9/8\,Ke^2\sim Ke^2$. \citep[see Appendix A in][]{lucy:2005}. 

\begin{figure}[htb]
\center
\includegraphics[angle=270, width=6in]{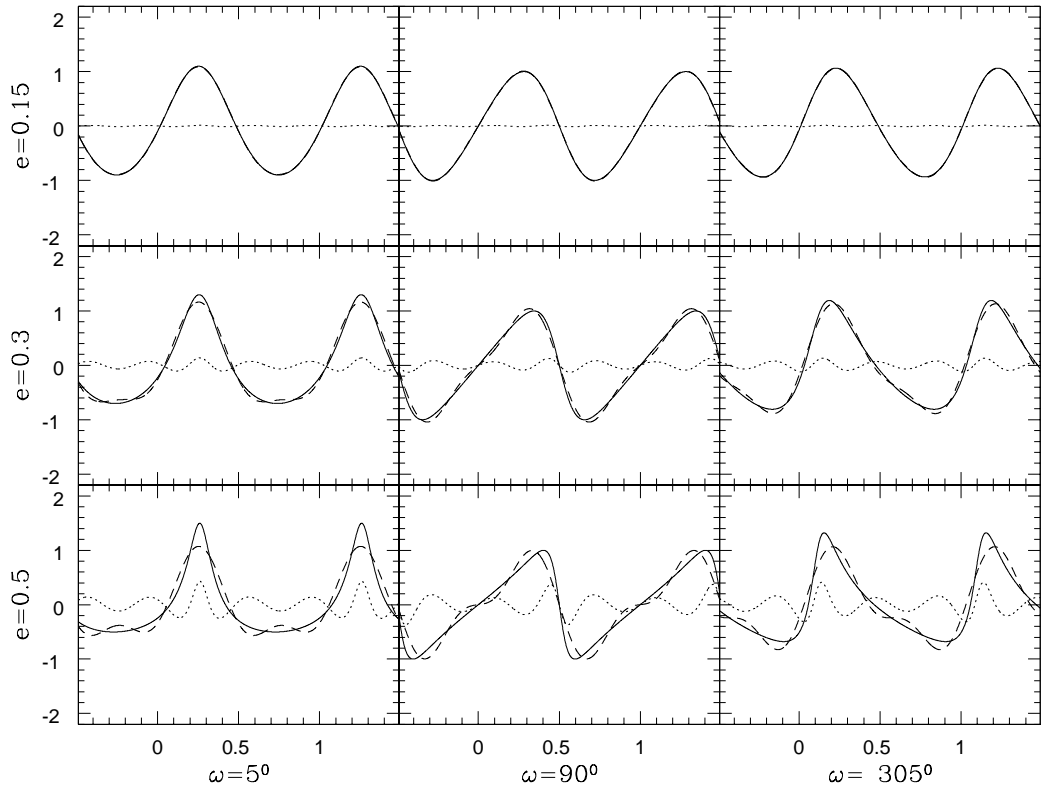}
\caption{
Radial velocity signals normalized to $K=1$. The eccentric
solution is the black solid line (eq.~\ref{eq:velecc}).  The
dashed line corresponds to the resonant case
(eq.~\ref{eq:velres}). The difference is clearly shown by as the
dotted line. In cases where $e<0.3$, both cases are harly
distinguishable (top panels).
}
\label{fig:resvsecc}
\end{figure}

The similarity of both solutions is illustrated in Figure
\ref{fig:resvsecc}, which shows how the Doppler radial velocity
curves would look like in each case (one-planet in an eccentric orbit
versus two resonant planets in circular orbits), for different values
of $\omega$ and $e$.  The two configurations can be easily confused,
especially when e $<$ 0.3 in the single planet case (or equivalently,
when the inner planet is significantly less massive than the outer
planet,i.e. $k_2 << k_1$). Confusion is also possible for larger
values of $e$ if the uncertainties are large and the radial velocity
curves are sparsely sampled, which is the case for several  published
Doppler velocity curves.  As an example estimate, if the detected
semi-amplitude and eccentricity are $K\sim 100~ms^{-1}$ and
$e\sim0.1$, the amplitude of the second harmonic will be $Ke^2\sim
1~ms^{-1}$ and both orbital solutions are indistinguishable at the
3--$\sigma$ level unless the precision of the data is better than
$0.3~ms^{-1}$. This is a problem, since only recently have planet
hunting groups started to achieve that level of precision
\citep{mayor:2008,fischer:2008}. All these statements will be made
more precise in Section \ref{sec:analysis}. There is also the
accuracy limitation imposed by the stellar jitter\footnote{intrinsic
noise associated to the stellar activity}, which has typical
amplitudes of 3--5 $ms^{-1}$ \citep{cumming:2008}. The optimal
strategies and the limitations of the Doppler technique to
disentangle this degeneracy are discussed in Sec.~\ref{sec:breaking}.

\section{Impact on known planetary systems}

Since, at least, 200 of the 350 known exoplanets have reported
eccentricities between 0.03 and 0.5, the mathematical degeneracy
described in the previous section may be affecting a large fraction
of known reported planets in eccentric orbits. Of course, this does
not imply that all the reported eccentric solutions must be resonant
pairs instead, but a singificant fraction of them may be, and this
can have a strong impact on our undersanding of formation of
planetary systems.

Assuming circular orbits, the mass of the inner companion
candidate $m_h$ can be estimated from the Keplerian solution as 
\begin{eqnarray}
m_{h} \sin i&=& \frac{e}{2^{1/3}}\,m_1 \sin i,
\end{eqnarray}

\noindent where $m_{1}$ is the mass of the outer planet and $e$ its
eccentricity, both parameters derived from the one--planet fit. The
$\sin i$ factor reflects the fact that the inclination $i$ (and the
true mass) is unknown when only Doppler information is available. When
a resonant two--planet system is confused with a single eccentric
planet, we refer to this situation as an \textit{eccentricity
imposter}. The term $Ke^2$ in Table 1 gives the amplitude of the
signal that needs to be resolved in all radial velocity candidates to
distinguish between solutions.

\subsection{Statistical analysis}\label{sec:analysis}

In order to quantify the extent of this degeneracy in the currently
known planet sample, we performed actual fits to most of the known
candidates with available data up to the date (March 2009, see
Exoplanet Encyclopedia$^2$). The data has been collected from two
sources, the systemic project web page
\textit{Systemic}\footnote{http://oklo.org, mantained by G.
Laughlin.} \citep{laughlin:2009}, and the NStED database
\footnote{http://nsted.ipac.caltech.edu/}. Still, there is a
significant number of published detections with no publicly available
datasets. The transiting exoplanets have been excluded from the list
because many of them have poorly sampled radial velocity curves and
their true eccentricity can be determined by other means such as the
photometric methods as described in  Sec.~\ref{breaking:photometry}.
Known multi-planetary systems have been excluded from the list as
well. These systems requires a more complex analysis
which addds unnecessary complications at this point. A number of
highly eccentric planets have not been added in the main sample 
because there is no reasonable doubt about their eccentric nature.
Their eccentricities, their corresponding $Ke^2$ and the SNR of the
second eccentric harmonic are given in Table \ref{tab:super} and they are
included in the statistical discussion at the end of this section.
The sample processed by our orbital fitting approach contains $163$
datasets listed in Table \ref{tab:statistics}.

Our method is based on a sequential fit of three different models:
circular, resonant and finally Keplerian. This approach takes
maximum advantage of the epicyclic decomposition of the radial
velocity signal as given in equation (\ref{eq:velecc}). For
practical purposes, equation (\ref{eq:velecc}) can be writen as
\begin{eqnarray}
v^e_r = \gamma &+& A \sin \left( Wt  \right) + 
          B \cos \left( W t \right) \label{eq:linearized} \\
      &+& C \sin \left( 2W t \right) + 
          D \cos \left( 2W t \right) \nonumber \\
      &+& \beta\, t\nonumber
\end{eqnarray}
\noindent which shows that the only severe non-linearity on the 
expression for the Doppler Keplerian signal is in the period. All
the other orbital parameters can be obtained as combinations of the
coefficients using basic trigonometric identities. The parameter
$\beta$ takes into account the signal of very long period objects
which appear as a linear trend. It is usually fitted to the published
solution and we will keep it as a free parameter in all that
follows. 

Step 1 consists on doing a linear Least Squares fitting(LS) of
$\gamma$, A, B and $\beta$ for many test periods. The best period is
the one which gives the minimum $\chi^2$. This is equivalent to the
classic Lomb--Scargle periodogram, but has the advantage that the
peaks (Least Squares minima in this case) and the coefficients from
the fit have a direct physical interpretation 
\citep[see][for a review on the topic]{cumming:2004}. 

In a second step, the data is fitted against the more complete model
containing $\gamma$, A, B, C, D and $\beta$  using many test periods
around the best circular solution found in step 1. Let us remark again
that the only strong nonlinearity lies on the period, so the fitting
of the linear parameters can be done very efficiently and without
danger of ending in a local minimum, a problem which plagues more
direct attack methods. The solution of this step gives the best
resonant orbital solution. 

Finally, the exact Keplerian expression for the radial velocity
\citep[see][]{lucy:1971} is fitted to the data using a non-linear
Least Squares approach. The seed values of the parameters for the
Keplerian solution are obtained from the resonant solution using the
map defined by equations (\ref{eq:velecc}) and (\ref{eq:velres}).
The optimal Keplerian fit is done using a straight-forward
non-linear LS minimization scheme using the analytic partial
derivatives of the Doppler signal with respect to the orbital
parameters \citep{numerical:1992}. The final Keplerian fits we
obtain are in good agreement with those found in the literature.
Therefore, as a by-product of this study, we also proof a powerful
method to attack the Kepler problem taking advantage of the
Linearized form of the Keplerian motion. The $\sqrt{\chi^2}$ of each
solution (circular, resonant and eccentric) are shown in columns 2,
3 and 4 on Table~\ref{tab:statistics}. 

The next step is to decide which of the orbital solutions is the
best (resonant or Keplerian) and whether it is significantly better
than the circular one. To do that, we apply the confidence level
test given by \citep{lucy:1971,lucy:2005}. We only accept one
of the non-circular solutions if the c.l. is better than $95\%$. 

In order to decide if the best non-circular solution (eg. resonant)
is statistically better than the other one(eg. Keplerian) we compute
the \textit{False Alarm Probability} of the favoured solution as
follows. Let us assume that the resonant solution is preferred, 
i.e.$\sqrt{\chi^2_{res}}<\sqrt{\chi^2_{ecc}}$. Then we generate a
synthetic data set using the best Keplerian solution. We add
Gaussian noise with a standard deviation equal to the RMS of the
Keplerian fit, and then obtain the best fit resonant configuration.
This process is repeated a large number of times. The number of
times when we get a $\sqrt{\chi^2}$ smaller than the real resonant
solution illustrates how an (un)fortunate combination of random
errors may be confusing a truly eccentric orbit with a resonant
one. On the contrary, if the eccentric solution is preferred, we
generate synthetic resonant data and fit for the Keplerian solution.
Since this is a computationally expensive process, the FAP is
initially computed based on $1000$ synthetic realizations of the
data. If the FAP is found to be smaller than $10\%$ the FAP is
recomputed using $10^5$ datasets. The result of this process is
illustrated in the last three columns of Table~\ref{tab:statistics},
which contains the preferred model, its FAP and a quality indicator:
* indicates a FAP$<5\%$, ** corresponds to FAP$<1\%$, *** is
FAP$<0.1\%$. If the solutions are not significantly different (FAP
$>5\%$) they are flaged as U (\textit{undecided}).

\begin{figure}[htb]
\center
\includegraphics[width=6in]{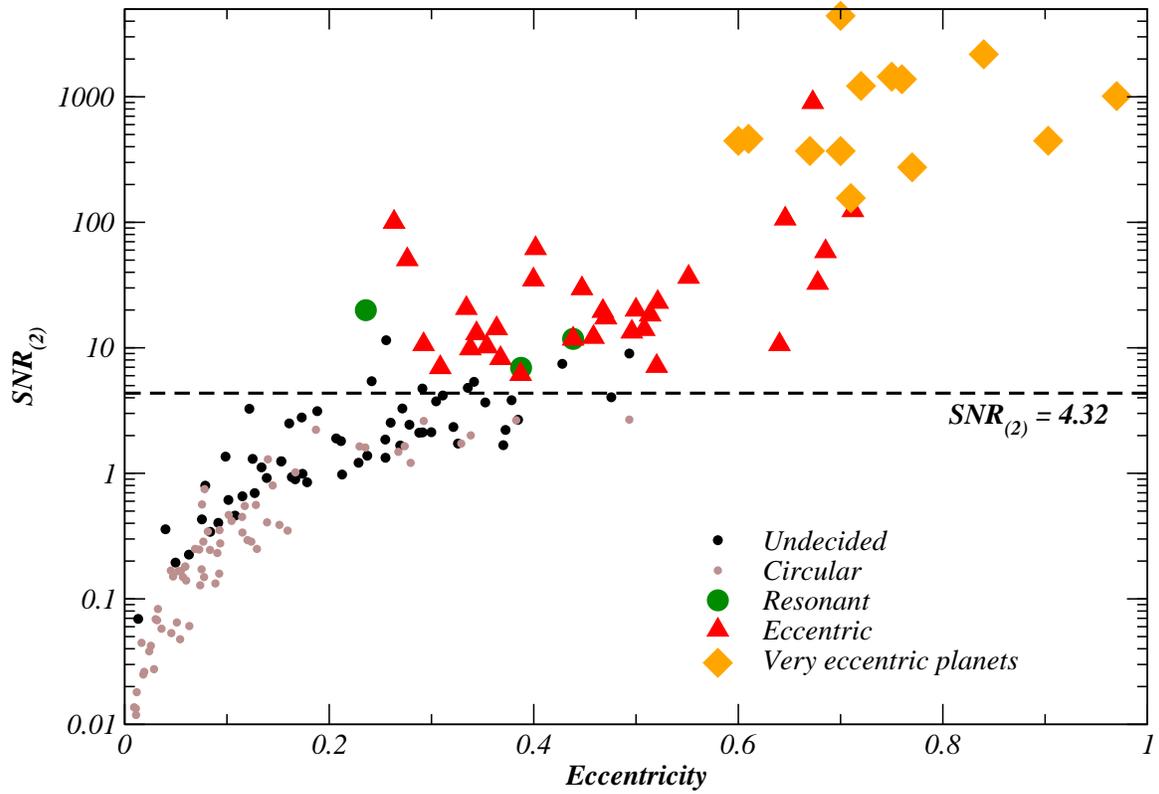}
\vspace{1.0in}
\caption{Orbital eccentricity versus the Signal-to-noise ratio of the 
$Ke^2$ harmonic. Only systems above the $4.32$ dashed line can be
statistically distinguished. The black dots are solutions where
the eccentric/resonant fit is significantly better than the circular one
but cannot be decided which solution is favoured
}\label{fig:evske2}
\end{figure}

The results of this procedure are also illustrated on
Fig.\ref{fig:evske2}. The quantity on the y--axis is defined as the
Signal-to-Noise ratio of the second harmonic and is computed from the
best Keplerian fit as 
\begin{eqnarray}
SNR_{(2)} =  \frac{K e^2}{RMS}\, \sqrt{N_{obs}}\label{eq:snr2}
\end{eqnarray}
\noindent  where $RMS$ is the root mean square of the residuals 
and $N_{obs}$ is the number of
observations. The horizontal line at $SNR_{(2)}=4.32$ is the minimal
theoretical threshold to detect the second Keplerian harmonic, assuming
that the required level of significance $p$ is $95\%$ as given in
\citet[][eq.~18]{lucy:2005}. The black points are the undecided ones,
this is, where the first eccentric harmonic is clearly significant
(circular solution discarded) but the statistics are insufficient to
decide which solution is significantly better (eccentric vs resonant). 

A word of caution has to be made here. The $\sqrt{\chi^2}$ are
obtained using the nominal uncertainities published with the
Doppler data. It is well established that most of the stars
introduce additional noise of astrophysical origin, that is usually
called \textit{stellar jitter}.  
The jitter makes it more difficult to disentangle the
degeneracy under discussion, meaning that more accurate
measurements with better spectrographs may not help. In the
randomly generated datasets, not considering the noise due to the
stellar jitter tends to give over-optimistic False Alarm
Probabilities. This is the reason why we use the RMS of
the solution rather than the published uncertainties 
to generate synthetic data.

In many cases, the current radial velocity data sets are sparsely
sampled, and more intensive monitoring at the most sensitive phases
is required (see Sec.~\ref{breaking:rvs}), especially for those
systems above the $4.32$ line of Fig.\ref{fig:evske2} and marked on
Tab.~\ref{tab:statistics} as undecided. We also show how the number
$Ke^2$ is a powerful discriminator to a priori decide if a dataset
is sensitive to the second eccentric harmonic, or equivalently, if
a dataset is able to disentangle the degeneracy under discussion by
just computing $SNR_{(2)}$ using the Keplerian parameters. 
In all the cases where an eccentric/resonant solution is
preferred against a circular one, we have added to column 6 in Table
\ref{tab:statistics} the number of required observations $N_{req}$ to
reach the $SNR_{(2)}$ of $4.32$ according to the eccentricity and RMS
of the best Keplerian solution using  equation \ref{eq:snr2}. A circular
orbit means that the first eccentric harmonic proportional to $Ke$ 
is already too small to be detected,so $N_{req}$ is not given.
In many
cases the number of required observations is extremly large. These
cases are indicated with a $+1000$. Even if the minimum number of
observations is reached, it is not always possible to distinguish
between solutions depending on the sampled phase. In such cases, a
few more points in the right phases should be enough to confirm a
resonant candidate or a given eccentricity. It is also important to
note that extending the time baseline can help disentagle
resonant systems if the periods are not exactly $2:1$ or if the
system is in a strong interacting regime. This point is discussed
again in Section \ref{sec:dynamics}.

Our results considering the sample of 163 planets plus the 13 very
eccentric ones in Table \ref{tab:super} are shown in Table \ref{tab:results}.
When the solution is clearly non-circular, our data processing
approach is unable to determine which solution is favoured in $63\%$
of the cases, which clearly proves the extent of the 
degeneracy. The statistical behaviour of the sample
is in good agreement with the predicted degeneracy threshold at
$SNR_{(2)}=4.32$. This gives us confidence in our data analysis
scheme and the method we propose to evaluate the statistical
significance of each solution using Monte Carlo generated False
Alarm Probabilities. 
A few systems with already known resonant
configurations which have gone to the process of : detect 
one eccentric planet and later discovery of a second planet in
a 2:1 configuration; are not considered in these counting
(eg. GJ 876 with 3 known planets, HD 128311 with 2 planets, 
HD 160691 with 4 planets). With this, we want to remark that 
highly ecentric candidates with unexplained large RMS (with large
$\chi^2$ in Tab.\ref{tab:statistics}) seem to be good 
targets to follow-up and uncover multiplanetary systems with very low 
mass companions.

\subsection{HD 125612b/c?}\label{sec:hd125612}

\begin{figure}[htb]
\center
\includegraphics[angle=0, width=7in, clip]{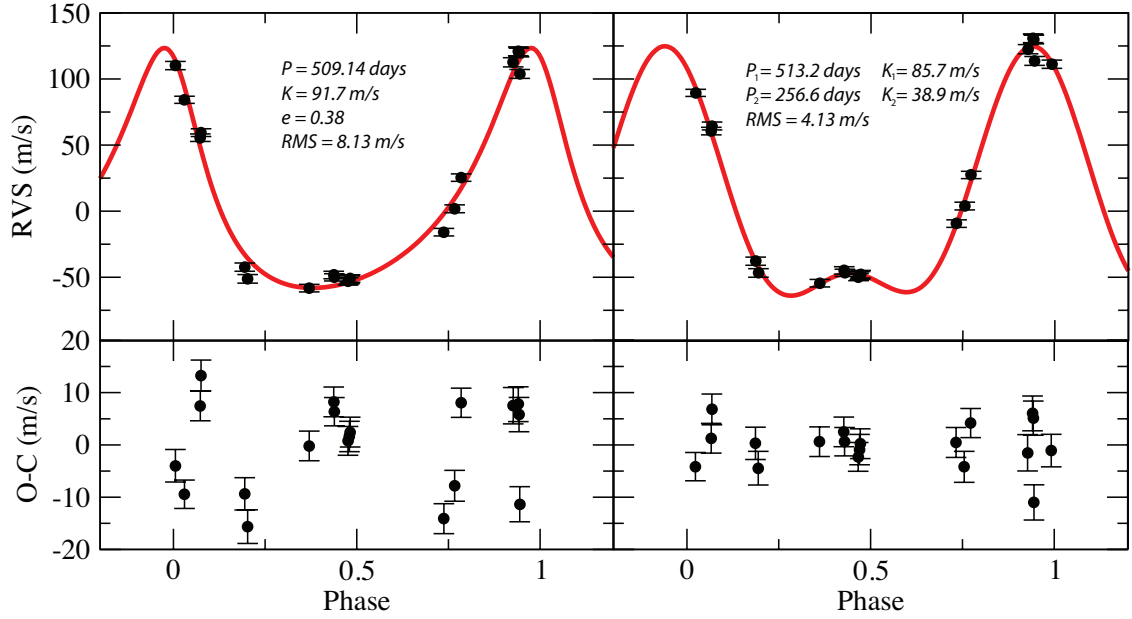}
\caption{Phased representation of the eccentric (left) and resonant
(right) solutions for system HD 125612. Assuming a stellar 
\textit{jitter} of 2.0 m s$^{-1}$), the square root of the reduced
$\chi^2$ is $\sqrt{\bar{\chi}^2}_{ecc}=3.58$ for the eccentric
solution and $\sqrt{\bar{\chi}^2}_{res}=1.64$ for the resonant one.
The residuals of each fit (bottom) clearly illustrate the reduction
of the dispersion in the resonant solution case.}
\label{fig:HD125612}
\end{figure}

A detailed analysis of our best resonant candidate system HD
125612, is discussed here to illustrate the procedure of
evaluating the  False Alarm probability in more detail. HD 125612b
is a gas giant planet detected around a nearby G3V star as
reported by \citet{fischer:2007} (from now on F07). A common
proper motion M4V companion has been recently associated to the
system \citet{mugrauer:2009}. The M4V is at a minimum distance of
$\sim 4750 AU$ from HD 125612A and has negligible effects on the
Doppler data during the time span of the observations. The star is
relatively quiet, so the expected jitter is of the order of $2.0$
m/s. F07 already pointed out an unexpectadly large RMS and
$\sqrt{\bar{\chi}^2}$, indicating the presence of additional
bodies in the system. We find that a resonant solution clearly
improves the quality of purely Keplerian fit
Fig.~\ref{fig:HD125612}). 
Assuming a stellar mass of $1.1 \,M_\odot$, the best-fit
masses of the putative resonant planets are $m_b = 3.2\pm0.4
\,m_{Jup}$ and $m_c = 1.1\pm0.3\, m_{Jup}$ with $P_b = 509 \pm15
$ days and $P_c=254.5$ days ($P_c$ is not a free parameter in
our resonant model). Since only $19$ data points are
available, we agree with F07 that more data is required to
disentangle the true nature of this system.

\begin{figure}[htb]
\center
\includegraphics[angle=0, width=5in]{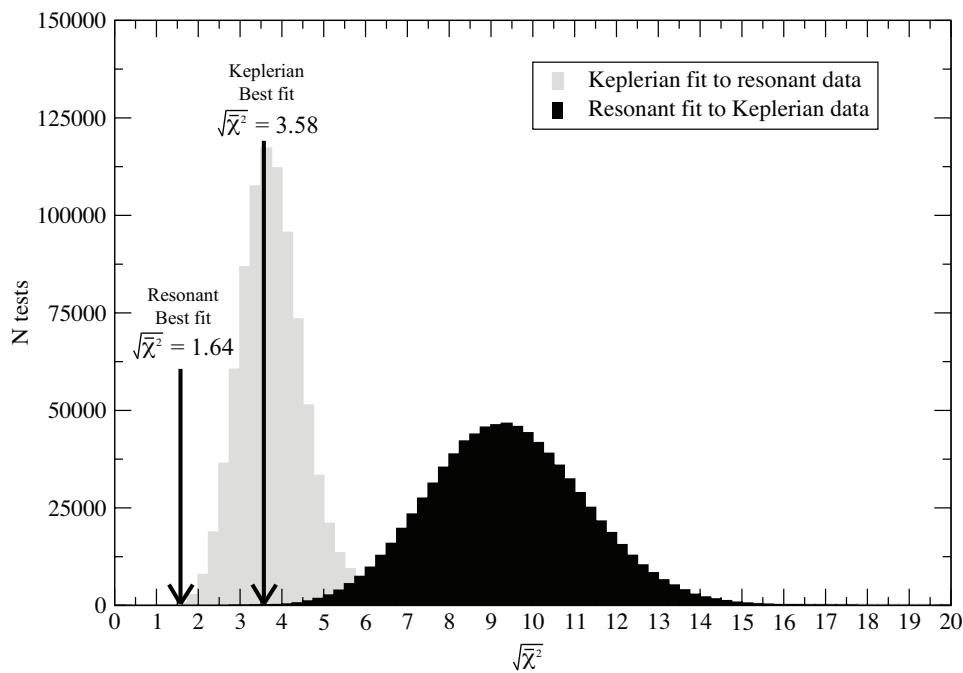}
\vspace{1.0in}
\caption{
Histograms of the Monte Carlo generated distributions
of$\sqrt{\bar{\chi}^2}$. Black bars are the result of fitting a resonant
model to the MC realizations  of the Keplerian solution. Our obtained
solution has a $\sqrt{\bar{\chi}^2}=1.64$ clearly smaller than the
typical  $\sqrt{\bar{\chi}^2}$ obtained adding noise to the Keplerian
model. Gray bars correspond to Keplerian fits to resonant simulated data.
This time the distribution peaks at $3.5$ very close to the best Keplerian
fit to the real data.
}
\label{fig:MC}
\end{figure}

To quantify how sigificant is the resonant solution with respect to
the eccentric one, we compute the empirical False Alarm Probability
as described in the previous section. For this experiment, we use
the square root of the reduced $\chi^2$, (i.e.
$\sqrt{\bar{\chi}^2}$) to enable direct comparison with F07.
$\sqrt{\bar{\chi}^2}$ differs from $\sqrt{\chi^2}$ used in Section
\ref{sec:analysis} by a constant multiplicative factor which is not
relevant for the FAP estimations because the number of free
parameters in the our resonant model is equal to the number of
parameters of a single planet Keplerian solution. In this case we
use the published uncertainities and the a nominal stellar jitter of
$2.0$ m/s to weight each observation and compute the
$\sqrt{\bar{\chi}^2}$. Since the FAP is very small, we produce one
million Monte Carlo realizations of the data. The histogram of the
obtained $\sqrt{\bar{\chi}^2}$ is shown in Fig.\ref{fig:MC}. In 350
out of one million cases, the resonant solution gives a better
$\sqrt{\bar{\chi}^2}$ than our best fit $1.64$, obtaining a false
alarm probability of $0.035\%$. 

We perform the same experiment but assuming an exact resonant
orbit, adding noise, and fitting a Keplerian orbit for another
set of one million MC realizations. In this case, the Monte Carlo
generated distribution of the $\sqrt{\bar{\chi}^2}$ peaks at $3.5$
very close to the $\sqrt{\bar{\chi}^2}$ published in F07. This
indicates that the obtained $\sqrt{\bar{\chi}^2} = 3.6$ by F07 is
compatible with the confusion of a resonant system with an
eccentric planet. We have repeated the experiment introducing
different levels noise (form the nominal $2.0$ m/s to  $10$ m/s),
obtaining very small FAP in all cases.

Therefore, we find strong statistical evidence to support the
hypothesis that HD 125612 has a pair of planets in the 2:1
resonance instead of a single eccentric planet. It clearly
exemplifies how the degeneracy under discussion and how the
natural bias to the eccentric orbital solutions can be affecting a
good number of exoplanet discoveries. 

\begin{figure}[htb]
\center
\includegraphics[angle=90, width=6in]{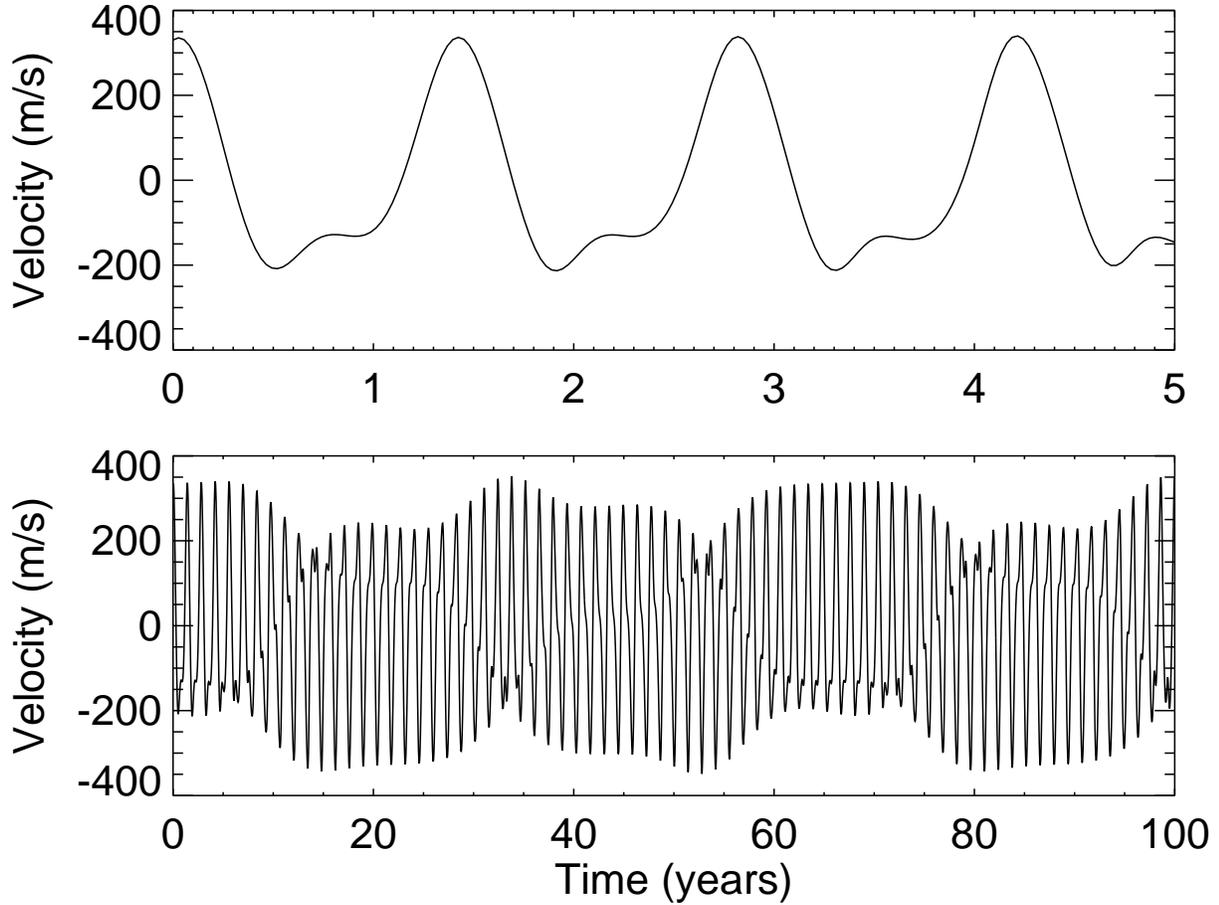}
\vspace{0.5in}
\caption{
Radial velocity signal as obtained by numerical integration of the
three body problem in HD 125612. Significant differences in the shape
of the signal start to be evident in time scales of $\sim$ 10 years.
}\label{hd125612}
\end{figure}

Finally, the dynamical stability of the system has been checked by
numerical orbital integrations up to 1 Myrs using Mercury
\citep{chambers:1999}. The integration of HD 125612 shows that the
candidate resonant planets would be in a strongly interacting regime
where one of the resonant critical arguments circulates and the
other one has a large libration amplitude. However, the system
remained stable and no close encounters occurred. Since the
orbital solution is poorly constrained, more data is required to
further constrain the orbital parameters and ensure the long term
stability of the system. The resulting Doppler signal as a
function of time is illustrated in Fig.~\ref{hd125612}. In the
Figure, it can be seen that the dynamincal interactions between 
planets will have obvious effects on the Doppler signal in
time-scales as short as a few orbital periods ($\sim 10$ years).

A small linear trend is required to obtain a good fit of the data
(in both Keplerian and resonant cases), which is too large to be
explained by the newly discovered M4V companion. This indicates
the presence of an additional very long period and massive planet
in the System. We strongly encourage the follow-up of this system.

\section{Dynamical stability}\label{sec:dynamics}

As illustrated in the discussion of HD125612b, a question that needs
to be addressed is whether the 2:1 resonant configurations in Table
1 are, in general, dynamically stable. Several 2:1 resonant
multi-planet systems have already been found (i.e. GJ 876, HD 82943,
HD 73526, HD 128311). Dynamical stability of 2:1 resonant
configurations has been also discussed by several authors. For
example, \cite{lee:2002} show that resonant locking may arise
naturally during the migration of exoplanets in the presence of a
protoplanetary disk.  The case where the outer planet is
significantly more massive has been recently discussed in great
detail by \cite{mitchenko:2008}, concluding that long term stability
is guaranteed and that the resonant capture during migration is
particularly favoured. Therefore, there is no theoretical objection
to the case addressed in this paper, on the contrary, recent work
strongly support the existence and stability of $k_{2} << k_{1}$
systems in 2:1 resonant configurations. 

Eccentric \textit{hot Neptunes} ($m_{1} \sin i< 50~m_{\earth}$),
are particularly interesting since their potentially hidden
companions are of a few Earth-masses or less (see  Table
\ref{tab:lowmass}). Their small $Ke^2$ makes it difficult to
distinguish between solutions with the current instrumental
accuracies but these are excellent targets to seek out for the
effects of dynamical interactions. A very tantalizing case is the
4 planet system around GJ 581 \citep{mayor:2009} where the inner
body is of a few earth masses and the new published solution for
the exterior planet (GJ 581d) gives an eccentricity of $0.4$. A
resonant planet hidden in the eccentric solution of GJ 581d would
perfectly fill the current gap between the 12 day orbit of GJ 581c
and 66 day orbit of GJ 581d. The potential candidate would be of a
few earth masses ($\sim 2-3 M_{\earth}$) and would lie in the
middle of the habitable zone.
Recent dynamical studies \citep{zollinger:2009} considering
all the planets except the small inner one, strongly support the
stability of the system when adding a few Earth-mass planet in a
fairly broad range of orbital configurations around the 2:1
resonance with GJ 581d. Since the inner planet is very small and
in a very tight orbit, the authors do not expect these results will
change too much. Given the RMS for GJ 581($\sim 1.46$ m/s), an
amplitude of the second harmonic for GJ 581d of $0.4$ m/s and
ignoring the dynamical interactions, the number of observations
required to reach a $SNR_{(2)}$ of $4.32$ and disentagle the 
degeneracy is $\sim 280$. Compared to the current $110$ observations,this
number could be achieved in near future.

\section{Breaking the degeneracy}\label{sec:breaking}

The question now becomes on how to observationally identify
\textit{eccentric imposters}. We focus here on the Doppler and
photometric methods, which are the only two techiques with
currently enough sensitivity to discern between both cases. In the
future, techniques  such as astrometry and direct imaging will be
useful as well.

\subsection{Using improved Doppler data}\label{breaking:rvs}

As discussed in previous sections, the most direct
approach is to increase the number of radial velocity observations,
$N_{obs}$, and their precision $\sigma_{obs}$, until the condition
$SNR_{(2)}>4.32$ is satisfied. It is important to recall that the
actual limit in pushing the accuracy $\sigma_{obs}$ is currently
put by stellar jitter that and can be of the order of $2-5~ms^{-1}$
even for relatively quiet stars. We suspect that some of the
undecided solutions in Fig \ref{fig:evske2} over the critical line
of $4.32$ are most likely dominated by stellar jitter or poorly
sampled making still undecidible which solution is favoured.

The phase of maximal difference will depend on each particular
combination of orbital parameters and has to be examined case by
case by direct inspection of the best Keplerian solution compared
to the best resonant one by subtracting both fitted models. As a
general rule, the differences will be more obvious near the
quadratures of the resonant solution, that is, around the extremes
of the doppler curve and at the quadratures of the inner candidate
seen as little bumps on the resonant signal (dashed line) in
Fig.~\ref{fig:resvsecc}.

This strategy works better when the initial one--planet fit
suggests a relatively large eccentricity $e>0.3$ and a poor
initial fit. An example of this situation was the discovery of the
planetary system around GJ 876. That system was initially confused
with a single planet \citep{delfosse:1998,marcy:1998} in an
$e\sim0.31$ orbit. Additional observations revealed a system of
two resonant planets with smaller eccentricities and in a strong
interaction regime \cite{marcy:2001}. Further observations and
detailed numerical integration of the N-body problem by
\cite{laughlin:2001} confirmed the presence of the two massive
eccentric bodies and uncovered an additional very short period
companion \citep{rivera:2005}.  

Therefore, in the case of planets in a strongly interacting regime 
(see  Sec.~\ref{sec:hd125612}), the numerical integration of the
orbits provides a powerful discrimination method and can be used
to predict the timescale required to observe the dynamical effects
which should be observed in the case of a resonant configuration.

\subsection{Photometric methods}\label{breaking:photometry}
A second approach is to use photometric observations. These can
confirm or discard the presence of a second planet in some
circumstances, either by detection of planetary transits and
occultations, or by observing reflected light or thermal emission
from the planets. Photometric methods are mostly efficient for
planets in short period orbits, since those tend to be hot and
have a higher probability of transiting in front of their star.
Assuming that the period $P$, the eccentricity $e$, and $\tau_0$
are known from the Doppler solution and that the orbital
inclination of the planet is close to $90^0$ (edge on), the
predicted instant of transit $T_I$ depends on the eccentricity as
\begin{equation}
T_{I}  = \tau_0 + NP\,\frac{1}{4} - 
\frac{eP\cos \omega}{\pi} + O({e^2}) , 
\end{equation}
\noindent where $N$ is the number of integer periods elapsed
from $\tau_0$. If the system contains a hidden companion, then
the true $e$ will be small and the transit will occur almost
exactly $1/4$P after crossing the line of nodes. This test is
only significant if $\tau_0$ is well constrained.  An
unambiguous determination of the eccentricity is obtained when
both the primary transit and the occultation (planet passes
behind the star), can be observed. This is because the time interval
between these two events is independent of $\tau_0$. If the
orbit is circular, the occultation occurs half period
after the transit. If the orbit of the planet is truly
eccentric, the time difference between the transit $T_I$ and
the occultation $T_{II}$ is
\begin{equation}
T_{II} - T_{I} = \frac{P}{2} + 2\,\frac{eP\cos\omega}{\pi} + O({e^2}),
\end{equation}
\noindent which can be as large as several hours on some of the
known transiting planets.

A representative example where the observation of transits brakes
the degeneracy is GJ 436. The star hosts a hot--neptune
in an eccentric orbit ($e\sim 0.14$) with a period of only $2.63$
days \citep{maness:2007}. The potential hidden companion would
have a mass as low as $2.5 m_{\earth}$. GJ 436b was recently found
to transit \citep{gillon:2007}, but because of the uncertainity in
$\tau_0$, the detection of the primary transit alone was not
sufficient to confirm the eccentric orbit. Shortly after, the
occultation was observed in thermal emission with the
\textit{Spitzer} telescope at the instant predicted by an
eccentric solution \citep{deming:2007}. If the orbit of GJ 436b
had been circular, the time of the occultation would differ by
three hours from the observed time and would not be detected with
the Spitzer observations.

\section{Conclusions}

We show that the Doppler signal of a single eccentric planet can
mimic the signal of a two-planet system in a 2:1 circular or
near--circular resonant orbit. This degeneracy is affecting a
large fraction of the known exoplanets, this is, around $30-40\%$
of the published single planet sytems. We also find strong
evidence of at least one case (HD 125612) where the resonant
solutions is significantly better than the published eccentric one
by \citet{fischer:2007}. The analysis described in this paper can
also be applied to multi-planet systems with eccentric candidates,
where the degree of degeneracy is expected to be similar or even
stronger due to the mixed signals of the different planets
involved. The detailed analysis of multiplanetary systems is more
complex and usually involves dynamical stability consideration.
Therefore, a case by case study is imposed. A remarkable example
is the planetary system around GJ 581 \citep{mayor:2009} were
$2-3$ earth mass planet could be hidden in the habitable zone of
the system.

The only techniques currently able to distinguish between two
resonant planets and single eccentric planet systems are limited
to the Doppler and photometric approaches described above. In the
future, other methods such as astrometry and direct imaging, will
also provide ways to uncover \textit{eccentric imposters}
\citep[see][]{moorhead:2009}. Astrometry will be the first one to
become sensitive enough, once the upcoming space astrometric
missions Gaia/ESA \citep[][to be launched in 2011]{lindegren:2008}
and SIM/NASA \citep[][to be launched after 2015]{unwin:2008}, go
on-line. High precision astrometry will be most helpful if the
resonant orbits are not coplanar. Otherwise, it will suffer from
the same degeneracies as the Doppler technique
\cite{konacki:2002}. The ultimate test will be direct imaging,
which will make possible to measure whether the orbit of the detected
planet is indeed eccentric. This will have to wait until
spaceborne missions such as a Darwin/TPF launch.

The conclusions of this work make it worth reconsidering some
published orbital solutions and motivate the follow-up of some
interesting systems. We find that future announcements of
eccentric planets should be carefully tested before publication
since it is relatively simple to check if there is any improvement
using a resonant configuration. Also, we find that several
reported radial velocity curves \citep{mayor:2008, mayor:2009} may
contain hidden signals of rocky planets.

\vspace{0.2in}

\textit{Acknowledgements.} GAE thanks A. Boss \& A. Weinberger for
financial support.  MLM acknowledges support provided by NASA
through Hubble Fellowship grant HF-01210.01-A awarded by the STScI,
which is operated by the AURA, Inc., for NASA, under contract
NAS5-26555. JEC would like to thank NASA's Origins of Solar Systems
Program for support. We also thank A. Bonanos and all the other
members of the Astronomy group at CIW-DTM for fruitful discussions.
The authors also recognize the comments and suggestions made by
L.Lucy and an anonymous referee which helped to improve the
manuscript significantly.


\newpage

\begin{deluxetable}{lcccc}
\tablecaption{
Historical evolution of the eccentricities in 55 Cnc. 
\label{55cnc}
}
\tablehead{
\colhead{Planet} &
\colhead{1997$^{a}$} & 
\colhead{2002$^{b}$} & 
\colhead{2004$^{c}$} & 
\colhead{2008$^{d}$}  \\
}
\startdata
e & --    & --    & 0.174 & 0.070 \\
b & 0.050 & 0.013 & 0.019 & 0.014 \\
c & --    & 0.080 & 0.440$^{c*}$ & 0.086 \\
f & --    & --    &    -- & 0.200   \\
d & --    & 0.146 & 0.327$^{c*}$ & 0.025 \\
\enddata
\tablenotetext{a}{\citet{butler:55cnc:1997}. Discovery paper}
\tablenotetext{b}{\citet{marcy:55cnc:2002}. A second planet is found.}
\tablenotetext{c}{\citet{mcarthur:55cnc:2004}.}
\tablenotetext{*}{A change of trend is seen here.
A different instrument and group was responsible for the discovery of the inner
2.8 days period and the new solution for the outer bodies
}
\tablenotetext{d}{\citet{fischer:2008}}

\end{deluxetable}

\begin{deluxetable}{lcccccccc}
\tablecaption{
List of planets with reported eccentricities. Data extracted from the
Extrasolar  Planet Encyclopedia (\texttt{http://exoplanet.eu}, mantained by 
J. Schneider). See the full table in the on--line material.\label{tab:lowmass}
}
\tablehead{
\colhead{Planet} & 
\colhead{$m \sin i $}  &
\colhead{P} &
\colhead{$e$} &
\colhead{$K$} &
\colhead{$Ke$} &
\colhead{$Ke^2$} &
\colhead{$m_{h} \sin i$} \\
\colhead{} & 
\colhead{($m_{\earth}$)}&
\colhead{(days)} &
\colhead{} &
\colhead{(m$s^{-1}$)} &
\colhead{(m$s^{-1}$)} &
\colhead{(m$s^{-1}$)} &
\colhead{($m_{\earth}$)} 
}
\startdata
GJ 581 c   &	 5.06 &    12.93 &     0.17 &	  3.38 &     0.58 &	0.10 &     0.68   \\
GJ 581 d   &	 6.69 &    66.80 &     0.38 &	  2.76 &     1.05 &	0.40 &     2.02   \\
HD 181433 b   &     7.14 &     9.37 &	  0.40 &     3.08 &	1.22 &     0.48 &     2.24   \\
HD 7924 b   &	  8.70 &     5.40 &	0.17 &     4.03 &     0.69 &	 0.12 &     1.17   \\
HD 69830 b   &     9.90 &     8.67 &	 0.10 &     3.80 &     0.38 &	  0.04 &     0.79   \\
HD 160691 c   &     9.96 &     9.64 &	  0.17 &     3.20 &	0.55 &     0.09 &     1.36   \\
55 Cnc e   &	10.20 &     2.82 &     0.07 &	  5.03 &     0.35 &	0.02 &     0.57   \\
GJ 674 b   &	11.10 &     4.69 &     0.20 &	  9.65 &     1.93 &	0.39 &     1.76   \\
HD 69830 c   &    11.40 &    31.56 &	 0.13 &     2.85 &     0.37 &	  0.05 &     1.18   \\
HD 190360 c   &    17.10 &    17.10 &     0.01 &     4.58 &	 0.05 &     0.00 &     0.14   \\
\ldots      &   \ldots &  \ldots  & \ldots   &  \ldots  &  \ldots  &  \ldots  &  \ldots     \\
\enddata
\end{deluxetable}

\begin{deluxetable}{lccccccrc}
\tablecaption{
Statistical comparison of circular, resonant and eccentric orbital solutions. The quality column highlights
the significance of the solution : *** is a secure solution, U indicates undecided. The estimated number
of required observations to detect the second harmonic $N_{req}$ is only given if the orbital solution
is significanly non-circular.\label{tab:statistics}}

\tablehead{
\colhead{Planet} & 
\colhead{ $\sqrt{\chi^2_{c}}$ } &
\colhead{ $\sqrt{\chi^2_{r}}$ } &
\colhead{ $\sqrt{\chi^2_{e}}$ } &
\colhead{NC c.l.} &
\colhead{$N_{obs}/N_{req}$} &
\colhead{Preferred} &
\colhead{FAP (\%)} &
\colhead{Quality} 
}
\startdata
6 Lyncis                        &    1.67 &    1.64 &	1.64 &  76.88 &   30 &  Circular    &	-          &        \\
14 And                          &    3.37 &    3.49 &	3.49 &   0.02 &   34 &  Circular    &	-          &        \\
14 Her                          &    3.43 &    1.86 &	1.60 &  99.99 &  119/30 &  Eccentric   &  10.00       &   U    \\
16 CygB                         &    3.14 &    2.58 &	1.22 &  99.99 &   95/1 &  Eccentric   &	0.20       &  **    \\
18 Del                          &    2.27 &    2.22 &	2.23 &  85.85 &   51 &  Circular    &	-          &        \\
42 Dra                          &    8.57 &    7.29 &	7.02 &  99.98 &   45/763 &  Eccentric   &  $\sim$50.0  &   U    \\
51 peg                          &    1.02 &    1.02 &	1.02 &  49.51 &  256 &  Circular    &	-          &        \\
70 vir                          &   14.94 &    6.93 &	1.33 &  99.99 &   74/1 &  Eccentric   &	0.00       &  ***   \\
81 Ceti                         &    2.10 &    1.76 &	1.69 &  99.89 &   33/361 &  Eccentric   &  $\sim$50.0  &   U    \\
$\beta$ Gem                     &    2.21 &    2.19 &	2.18 &  86.34 &   80 &  Circular    &	-          &        \\
BD-10 3166                      &    1.55 &    1.58 &	1.58 &  39.39 &   31 &  Circular    &	-          &        \\
ChaHa8                          &    0.92 &    0.89 &	0.88 &  76.72 &   15 &  Circular    &	-          &        \\
$\epsilon$ Eri                  &    2.29 &    2.29 &	2.25 &  96.03 &  120/150\tablenotemark{a} &  Eccentric   &	0.10       &  ***   \\
GJ 176                          &    3.58 &    3.56 &	3.48 &  90.57 &   57 &  Circular    &	-          &        \\
GJ 3021                         &    5.66 &    3.49 &	1.86 &  99.99 &   61/21 &  Eccentric   &	0.00       &  ***   \\
GJ 849                          &    1.62 &    1.62 &	1.62 &  64.53 &   29 &  Circular    &	-          &        \\
HD 142                          &    1.11 &    0.81 &	0.79 &  99.97 &   27/47 &  Eccentric   &  50.00       &   U    \\
HD 2638                         &    2.23 &    2.30 &	2.30 &  27.13 &   28 &  Circular    &	-          &        \\
HD 3651                         &    2.54 &    2.12 &	2.04 &  99.99 &  121/10 &  Eccentric   &  4.00       &   *    \\
HD 4203                         &    4.77 &    2.39 &	1.29 &  99.99 &   23/1 &  Eccentric   &	0.00       &  ***   \\
HD 4208                         &    1.17 &    1.17 &	1.18 &  69.01 &   41 &  Circular    &	-          &        \\
HD 4308                         &    1.36 &    1.34 &	1.35 &  78.26 &   41 &  Circular    &	-          &        \\
HD 5319                         &    3.36 &    3.33 &	3.35 &  70.49 &   30 &  Circular    &	-          &        \\
HD 6434                         &    1.52 &    1.40 &	1.41 &  99.99 &  130/+1000 &  Resonant    &	6.50       &  U     \\
HD 7924                         &    3.81 &    3.77 &	3.78 &  83.83 &   93 &  Circular    &	-          &        \\
HD 8574                         &    1.69 &    1.32 &	1.13 &  99.99 &   41/148 &  Eccentric   &  $\sim$50.00 &   U    \\
HD 10647                        &    1.56 &    1.53 &	1.56 &  87.85 &   70 &  Circular    &	-          &        \\
HD 10697                        &    3.99 &    2.39 &	2.48 &  99.99 &   59/415 &  Resonant    &  $\sim$50.00 &   U    \\
HD 11977                        &    1.43 &    1.15 &	1.15 &  99.99 &   42/+1000 &  Resonant    &  34.0        &  U   \\
HD 12661                        &    2.91 &    2.02 &	2.13 &  99.99 &   51/+1000 &  Resonant    &  15.00       &   U    \\
HD 13189                        &   12.01 &    9.79 &  10.23 &  99.99 &   91/+1000 &  Resonant    &	3.30       &  *     \\
HD 13445                        &    3.56 &    2.67 &	2.68 &  99.99 &   42/+1000 &  Resonant    &  $\sim$50.00 &   U    \\
HD 14810                        &   45.96 &   24.65 &  22.13 &  99.99 &   30/+1000 &  Eccentric   &  $\sim$50.00 &   U    \\
HD 16141                        &    1.64 &    1.49 &	1.47 &  99.96 &   71/+1000 &  Eccentric   &  $\sim$50.00 &   U    \\
HD 16417                        &    3.73 &    3.50 &	3.53 &  99.77 &   88 &  Resonant    &  $\sim$50.00 &   U    \\
HD 19994                        &    1.88 &    1.56 &	1.56 &  99.98 &   48/458 &  Eccentric   &  19.00       &   U    \\
HD 20367                        &    1.29 &    1.19 &	1.24 &  93.49 &   27 &  Circular    &	-          &        \\
HD 17092                        &    3.11 &    2.95 &	2.95 &  97.58 &   59/+1000 &  Eccentric   &  $\sim$50.00 &   U    \\
HD 23079                        &    1.05 &    0.82 &	0.90 &  98.31 &   19/813 &  Resonant    &  $\sim$50.00 &   U    \\
HD 23127                        &    3.54 &    3.14 &	2.85 &  99.91 &   34/152 &  Eccentric   &  29.00       &   U    \\
HD 24040                        &    3.64 &    3.56 &	3.55 &  77.40 &   26 &  Circular    &	-          &        \\
HD 27442                        &    4.22 &    3.97 &	3.96 &  98.39 &   55/+1000 &  Eccentric   &  $\sim$50.00 &   U    \\
HD 27894                        &    2.57 &    2.49 &	2.47 &  78.78 &   20 &  Circular    &	-          &        \\
HD 28185                        &    1.67 &    1.60 &	1.59 &  92.89 &   40 &  Circular    &	-          &        \\
HD 28185                        &    1.44 &    1.51 &	1.51 &  42.72 &   15 &  Circular    &	-          &        \\
HD 28305                        &    2.71 &    2.26 &	2.17 &  98.40 &   20/800 &  Eccentric   &  $\sim$50.00 &   U    \\
HD 30177                        &    1.31 &    1.16 &	1.15 &  88.35 &   15 &  Circular    &	-          &        \\
HD 33283                        &    1.44 &    0.82 &	0.76 &  99.99 &   25/39 &  Eccentric   &  40.00 &   U    \\
HD 33636$^{c}$                  &    8.84 &    2.46 &	1.41 &  99.99 &   21/1 &  Eccentric   &	0.00       &  ***   \\
HD 39091                        &   19.40 &   10.88 &	1.29 &  99.99 &   42/1 &  Eccentric   &	0.00       &  ***   \\
HD 81688                        &    4.00 &    3.98 &	3.75 &  93.74 &   34 &  Circular    &	-          &        \\
HD 88133                        &    1.57 &    1.43 &	1.51 &  86.87 &   17 &  Circular    &	-          &        \\
HD 40979                        &    2.69 &    2.04 &	1.96 &  99.99 &   39/+1000 &  Eccentric   &  $\sim$50.00 &   U    \\
HD 41004B                       &    1.12 &    1.01 &	0.99 &  99.99 &  149/+1000 &  Eccentric   &  $\sim$50.00 &   U    \\
HD 43691                        &    1.28 &    1.26 &	1.27 &  70.36 &   22 &  Circular    &	-          &        \\
HD 43848                        &    7.68 &    0.57 &	1.57 &  99.99 &   10/1\tablenotemark{b}    &  Resonant    &	0.00       &  ***   \\
HD 46375                        &    1.77 &    1.72 &	1.72 &  88.86 &   50 &  Circular    &	-          &        \\
HD 48265                        &    1.53 &    1.47 &	1.49 &  74.82 &   17 &  Circular    &	-          &        \\
HD 49674                        &    1.56 &    1.60 &	1.60 &  26.89 &   39 &  Circular    &	-          &        \\
HD 50499d                       &    1.03 &    1.06 &	1.06 &  31.51 &   28 &  Circular    &	-          &        \\
HD 50554                        &    3.75 &    2.42 &	1.31 &  99.99 &   40/20 &  Eccentric   &	3.90       &  *     \\
HD 52265                        &    1.47 &    1.15 &	1.14 &  99.99 &   91/117 &  Eccentric   &  26.00       &   U    \\
HD 63454                        &    2.85 &    2.62 &	2.59 &  94.72 &   26 &  Circular    &	-          &        \\
HD 64468                        &  266.72 &   30.94 &	4.28 &  99.99 &   13/1 &  Eccentric   &	0.00       &  ***   \\
HD 65216                        &    1.63 &    1.27 &	1.22 &  99.99 &   70/40 &  Eccentric   &  15.00 &   U    \\
HD 66428                        &    4.30 &    2.08 &	1.10 &  99.99 &   29/1 &  Eccentric   &	4.80       &  *     \\
HD 68988                        &    7.00 &    4.54 &	4.48 &  99.99 &   28/+1000 &  Eccentric   &  $\sim$50.00 &   U    \\
HD 70573                        &    2.02 &    2.06 &	2.03 &  54.80 &   34 &  Circular    &	-          &        \\
HD 70642                        &    1.30 &    1.31 &	1.31 &  59.61 &   28 &  Circular    &	-          &        \\
HD 72659                        &    1.87 &    1.85 &	0.00 &  59.61 &   28 &  Circular    &	-          &        \\
HD 73108                        &   12.46 &    6.46 &	5.46 &  99.99 &   59/104 &  Eccentric   &	7.50       &  U     \\
HD 73267                        &    5.25 &    1.40 &	1.15 &  99.99 &   39/1 &  Eccentric   &  0.01        &  ***    \\
HD 75289                        &    0.86 &    0.86 &	0.86 &  52.20 &   88 &  Circular    &	-          &        \\
HD 76700                        &    1.33 &    1.28 &	1.28 &  85.94 &   35 &  Circular    &	-          &        \\
HD 81040                        &    2.70 &    2.53 &	2.00 &  99.91 &   26/115 &  Eccentric   &  $\sim$50.00 &   U    \\
HD 83443                        &    1.64 &    1.64 &	1.64 &  31.80 &  257 &  Circular    &	-          &        \\
HD 86081                        &    1.14 &    1.03 &	1.03 &  95.11 &   26/+1000 &  Resonant    &  $\sim$50.00 &   U    \\
HD 88133                        &    1.98 &    1.89 &	1.96 &  81.68 &   21 &  Circular    &	-          &        \\
HD 89307                        &    0.62 &    0.52 &	0.55 &  86.41 &   12 &  Circular    &	-          &        \\
HD 89744                        &    8.57 &    6.57 &	1.44 &  99.99 &   85/3 &  Eccentric   &	0.00       &  ***   \\
HD 92788                        &    3.47 &    2.27 &	1.63 &  99.99 &   55/25 &  Eccentric   &	3.60       &  *     \\
HD 93083                        &    2.15 &    1.76 &	1.75 &  95.28 &   16/47 &  Eccentric   &  $\sim$50.00 &   U    \\
HD 99109                        &    1.62 &    1.66 &	1.66 &  17.54 &   41 &  Circular    &	-          &        \\
HD 99492                        &    1.43 &    1.32 &	1.34 &  98.69 &   51/122 &  Resonant    &  $\sim$50.00 &   U    \\
HD 100777                       &    4.06 &    1.71 &	1.22 &  99.99 &   29/5 &  Eccentric   &  10.00       &  U     \\
HD 101930                       &    2.09 &    2.17 &	2.20 &  45.74 &   16 &  Circular    &	-          &        \\
HD 102117                       &    1.05 &    1.05 &	1.05 &  55.34 &   44 &  Circular    &	-          &        \\
HD 102195                       &    1.18 &    1.18 &	1.34 &  65.49 &   21 &  Circular    &	-          &        \\
HD 104985                       &    4.32 &    4.15 &	4.12 &  95.85 &   52/+1000 &  Eccentric   &  $\sim$50.00 &   U    \\
HD 106252                       &    4.41 &    1.98 &	1.11 &  99.99 &   40/9 &  Eccentric   &	0.00       &  ***   \\
HD 107148                       &    1.47 &    1.51 &	1.52 &  20.65 &   35 &  Circular    &	-          &        \\
HD 108147                       &    2.07 &    1.60 &	1.33 &  99.99 &  118/80 &  Eccentric   &	0.00       &  ***   \\
HD 109749                       &    1.08 &    1.04 &	1.04 &  78.12 &   21 &  Circular    &	-          &        \\
HD 114386                       &    1.74 &    1.50 &	1.59 &  99.99 &   58/+1000 &  Resonant    &  40.00       &   U    \\
HD 114729                       &    1.63 &    1.58 &	1.53 &  96.33 &   42/82 &  Eccentric   &  42.00 &   U    \\
HD 114762                       &    6.90 &    2.88 &	1.08 &  99.99 &   45/23 &  Eccentric   &	0.00       &  ***   \\
HD 114783                       &    1.82 &    1.77 &	1.77 &  89.45 &   54 &  Circular    &	-          &        \\
HD 117207                       &    1.38 &    1.23 &	1.25 &  99.35 &   43/739 &  Resonant    &  27.00       &   U    \\
HD 117618                       &    1.43 &    1.41 &	1.35 &  97.93 &   57/73 &  Eccentric   &  30.00 &   U    \\
HD 118203                       &    3.60 &    1.90 &	1.50 &  99.99 &   43/147 &  Eccentric   &  17.00       &   U    \\
HD 121504                       &    1.92 &    1.93 &	1.93 &  52.03 &  100 &  Circular    &	-          &        \\
HD 125612                       &   13.77 &    2.16 &	4.82 &  99.99 &   19/21 &  Resonant    &	0.03       &  ***   \\
HD 130322                       &    1.33 &    1.29 &	1.29 &  98.85 &  118/+1000 &  Eccentric   &  30.00       &   U    \\
HD 134987                       &    3.60 &    1.83 &	1.68 &  99.99 &   56/30 &  Eccentric   &  15.00 &   U    \\
HD 136118                       &    2.96 &    1.48 &	1.25 &  99.99 &   37/101 &  Eccentric   &	3.40       &  *     \\
HD 139357                       &    1.91 &    1.75 &	1.75 &  99.20 &   49/+1000 &  Resonant    &  45.00       &   U    \\
HD 141937                       &    4.06 &    2.29 &	1.58 &  99.99 &   81/2 &  Eccentric   &	0.10       &  ***   \\
HD 142022A                      &    2.40 &    1.57 &	1.37 &  99.99 &   49/13 &  Eccentric   &	0.80       &  **    \\
HD 142091                       &    2.24 &    2.28 &	2.26 &  53.81 &   46 &  Circular    &	-          &        \\
HD 143361                       &    1.95 &    0.85 &	1.24 &  97.54 &   12/13 &  Resonant   &   8.00 &   U    \\
HD 145377                       &   26.00 &   10.87 &	9.02 &  99.99 &   64/50 &  Eccentric   &  40.00 &   U    \\
HD 147513                       &    1.88 &    1.63 &	1.59 &  99.35 &   30/349 &  Eccentric   &  33.00       &   U    \\
HD 149026                       &    3.11 &    3.18 &	3.17 &  40.26 &   30 &  Circular    &	-          &        \\
HD 149143                       &    1.16 &    1.15 &	1.15 &  67.07 &   17 &  Circular    &	-          &        \\
HD 150706                       &    0.98 &    0.82 &	0.76 &  98.47 &   19/183 &  Eccentric   &  $\sim$50.00 &   U    \\
HD 153950                       &    6.67 &    3.03 &	2.42 &  99.99 &   49/20 &  Eccentric   &  40.00 &   U    \\
HD 154345                       &    2.77 &    2.70 &	2.65 &  95.43 &   55/194 &  Eccentric   &  $\sim$50.00 &   U    \\
HD 154672                       &    8.68 &    5.06 &	1.80 &  99.99 &   16/1 &  Eccentric   &	0.01       &  ***    \\
HD 154857                       &   11.39 &    8.19 &	8.54 &  99.97 &   28/726 &  Resonant    &  $\sim$50.00 &   U    \\
HD 160691                       &    5.78 &    4.45 &	4.25 &  99.99 &  108/46 &  Eccentric   &	1.10       &  *     \\
HD 162020                       &   32.32 &    7.82 &	1.61 &  99.99 &   46/2 &  Eccentric   &	0.00       &  ***   \\
HD 164922                       &    1.57 &    1.59 &	1.59 &  22.99 &   64 &  Circular    &	-          &        \\
HD 167042                       &    1.51 &    1.57 &	1.57 &   1.76 &   31 &  Circular    &	-          &        \\
HD 167042                       &    1.28 &    1.33 &	1.33 &  14.13 &   29 &  Circular    &	-          &        \\
HD 168746                       &    1.50 &    1.50 &	1.50 &  79.29 &  154 &  Circular    &	-          &        \\
HD 169822                       &   20.44 &    5.64 &	1.67 &  99.99 &   21/1 &  Eccentric   &	0.00       &  ***   \\
HD 170469                       &    2.38 &    2.44 &	2.43 &  28.36 &   35 &  Circular    &	-          &        \\
HD 173416                       &    1.71 &    1.50 &	1.50 &  99.90 &   52/+1000 &  Resonant    &  $\sim50.00$       &  U     \\
HD 175541                       &    2.80 &    2.66 &	2.57 &  94.80 &   29 &  Circular    &	-          &        \\
HD 177830                       &    3.21 &    3.14 &	3.14 &  87.75 &   54 &  Circular    &	-          &        \\
HD 178911B                      &    2.76 &    1.13 &	1.06 &  99.99 &   44/187 &  Eccentric   &  $\sim$50.00 &   U    \\
HD 179949                       &    2.16 &    2.15 &	2.15 &  72.02 &   65 &  Circular    &	-          &        \\
HD 185269                       &    3.08 &    2.07 &	1.74 &  99.99 &   30/68 &  Eccentric   &  $\sim$50.00 &   U    \\
HD 188015                       &    1.80 &    1.44 &	1.44 &  99.99 &   44/248 &  Eccentric   &  $\sim$50.00 &   U    \\
HD 189733                       &    7.33 &    6.15 &	6.12 &  93.88 &   16 &  Circular    &	-          &        \\
HD 190228                       &    2.84 &    1.36 &	0.98 &  99.99 &   51/4 &  Eccentric   &	0.10       &  ***   \\
HD 190647                       &    2.41 &    1.00 &	1.24 &  99.99 &   21/9 &  Resonant    &  10.00 &   U    \\
HD 192263                       &    1.64 &    1.65 &	1.65 &  43.98 &  181 &  Circular    &	-          &        \\
HD 192699                       &    1.97 &    1.90 &	1.89 &  89.53 &   34 &  Circular    &	-          &        \\
HD 195019                       &    1.66 &    1.67 &	1.67 &  42.31 &  117 &  Circular    &	-          &        \\
HD 196050                       &    2.62 &    1.79 &	1.65 &  99.99 &   44/103 &  Eccentric   &  29.00       &   U    \\
HD 205739                       &    2.99 &    2.51 &	2.83 &  98.45 &   24/348 &  Resonant    &  $\sim$50.00 &   U    \\
HD 208487                       &    1.63 &    1.49 &	1.47 &  98.08 &   35/356 &  Eccentric   &  $\sim$50.00 &   U    \\
HD 209458                       &    1.60 &    1.61 &	1.61 &   6.39 &  141 &  Circular    &	-          &        \\
HD 210277                       &    5.03 &    3.07 &	1.61 &  99.99 &   69/1 &  Eccentric   &	0.00       &  ***   \\
HD 210702                       &    1.64 &    1.61 &	1.59 &  80.38 &   29 &  Circular    &	-          &        \\
HD 212301                       &    2.39 &    2.14 &	2.11 &  95.41 &   23/+1000 &  Eccentric   &  $\sim$50.00 &   U    \\
HD 213240                       &    3.10 &    1.78 &	1.84 &  99.99 &   72/30 &  Resonant    &	2.80       &  *     \\
HD 216435                       &    1.71 &    1.74 &	1.74 &  14.41 &   58 &  Circular    &	-          &        \\
HD 216437                       &    3.23 &    1.61 &	1.60 &  99.99 &   39/25 &  Eccentric   &  39.00       &   U    \\
HD 216770                       &    1.32 &    1.23 &	1.11 &  93.40 &   16 &  Circular    &	-          &        \\
HD 224693                       &    1.27 &    1.31 &	1.30 &  47.66 &   24 &  Circular    &	-          &        \\
HD 231701                       &    1.84 &    1.87 &	1.84 &  63.22 &   17 &  Circular    &	-          &        \\
HD 330075                       &    1.43 &    1.52 &	1.52 &   7.73 &   21 &  Circular    &	-          &        \\
HIP 75458                       &   27.84 &   21.44 &	3.15 &  99.99 &   119/1 &  Eccentric   &	0.00       &  ***   \\
HR 810                          &    1.68 &    1.56 &	1.54 &  99.99 &   95/+1000 &  Eccentric   &  40.00       &  U     \\
ksi Aquila                      &    3.85 &    3.71 &	3.58 &  91.42 &   26 &  Circular    &	-          &        \\
NGC 2423 3                      &    2.76 &    2.48 &	2.41 &  98.00 &   28/+1000 &  Eccentric   &  $\sim$50.00 &   U    \\
NGC 4349 127                    &    6.75 &    6.45 &	6.23 &  88.14 &   20 &  Circular    &	-          &        \\
$\rho$ CrB                      &    1.11 &    1.08 &	1.07 &  97.24 &   79/+1000 &  Eccentric   &  27.00       &   U    \\
$\tau$ Boo                      &    1.92 &    1.91 &	1.91 &  65.51 &   98 &  Circular    &	-          &        \\
\enddata
\tablenotetext{a}{Other data sets and astrometry seems to confirm the eccentricity. Very noisy star}
\tablenotetext{b}{The obtained solution differs from the published one significantly}
\tablenotetext{c}{According to \citep{Bean:2007}, astrometric observations
indicate that the candidate is a star indeed with an
orbital inclination close to $~0$ }
\end{deluxetable}

\begin{deluxetable}{lccc}
\tablecaption{Already known eccentric planets not included in
Table \ref{tab:statistics}.\label{tab:super} }
\tablehead{
\colhead{Planet} & 
\colhead{e} & 
\colhead{$Ke^2$} &
\colhead{$SNR_{(2)} $} 
}
\startdata
HD 4113   & 0.903 & 335 & 446 \\
HD 156846 & 0.84 & 2100 & 2181 \\
HD 20782 & 0.97 & 438 & 1012 \\
HD 222782 & 0.76 & 924 & 1380 \\
HD 20868 & 0.75 & 355 & 1446 \\
HD 75458 & 0.72 & 1496 & 1218\\
HD 96167 & 0.71 & 115& 156 \\
HD 159868 & 0.7 & 279 & 369 \\
HD 2039 & 0.67 &  785  & 370   \\
HD 37605 & 0.77 & 370  & 274  \\
HD 131664 & 0.7 & 2757  &  4413 \\
HD 171028 & 0.61 & 618  & 462   \\
HD 16175 & 0.6 & 618 & 445  \\
\enddata
\end{deluxetable}

\begin{deluxetable}{lcc}
\tablecaption{Statistical results from our sample of exoplanets.\label{tab:results}}
\tablehead{
\colhead{Solution type} & 
\colhead{Number of cases} &
\colhead{Fraction} 
}
\startdata
Total        & 176 & $ 100\%$ \\
\\
Circular     &  71 & $ 40\%$ \\
Non-circular & 105 & $ 60\%$ \\
\\
Eccentric    &  38 & $ 22\%$ \\
Resonant     &   4 & $ \sim 2\%$ \\
Undecided    &  63 & $ 36\%$ \\
\enddata
\end{deluxetable}

\end{document}